\documentclass[%
 reprint, showkeys,
 amsmath,amssymb,
 aps,
]{revtex4-2}

\usepackage{graphicx}% Include figure files
\usepackage{dcolumn}% Align table columns on decimal point
\usepackage{bm}% bold math
\begin{document}

%\preprint{APS/123-QED}

\title{On the relativistic viability of multi-automaton systems: \\
essential concepts, challenges and prospects}

\author{Alexandru-Ionu\c{t} B\u{a}beanu}
 \email{a.i.babeanu@gmail.com}
\affiliation{% Authors' institution and/or address\\
 Delft University of Technology \\
 Rotterdam School of Management, Erasmus University
}%

\date{\today}% It is always \today, today,

\begin{abstract}
Our understanding of the Universe breaks down for very small spacetime intervals, corresponding to an extremely high level of granularity (and energy), commonly referred to as the ``Planck scale''. At this fundamental level, there are attempts of describing physics in terms of interacting automata that perform classical, deterministic computation. On one hand, various mathematical arguments have already illustrated how quantum laws (which describe elementary particles and interactions) could in principle arise as low-granularity approximations of automata-based systems. On the other hand, understanding how such systems might give rise to relativistic laws (which describe spacetime and gravity) remains a major problem. I explain here a few ideas that seem crucial for overcoming this problem, along with related algorithmic challenges that need to be addressed. Giving emphasis to meaningful computational counterparts of locality and general covariance, I outline basic ingredients of a distributed communication-rewiring protocol that would allow us to construct multi-automaton models that are viable from a relativistic perspective. I also explain how viable models can be evaluated using a variety of criteria, and discuss related aspects pertaining to the falsifiability and plausibility of the automata paradigm.

\end{abstract}

\keywords{Strict locality, distributed computation, causal sets, asynchronous communication, pregeometry, multi-agent systems, superdeterminism, automata models.}%Use showkeys class option if keyword
\maketitle

This work is motivated by an outstanding challenge \cite{Carlip2001} in physics: understanding how the Universe operates at its most fundamental level, known as the Planck scale. This is associated to very high spacetime granularities, well beyond the reach of the most powerful experiments for particle physics, so theoretical research is crucial for gaining insights. The challenge is to combine two existing theories, which are empirically very successful at lower granularities: quantum theory (describing elementary particles and interactions) and the theory of general relativity (describing spacetime and gravity). While essential aspects of both theories are expected to play important roles at the Planck scale, integrating them in a self-consistent and falsifiable way remains a serious problem \cite{Smolin2018}, despite extensive efforts that go back one century. Besides potentially telling us what the world (including spacetime itself) is really made of, a Planck scale theory is crucial for understanding the very early Universe and the cores of black holes. It should also provide insights about the mechanism responsible for gravitational anomalies commonly attributed to dark matter \cite{Bertone2018}, and the feasibility of spacetime manipulations, like the Alcubierre drive \cite{Alcubierre1994}, that could facilitate interstellar travel.

Almost all theoretical efforts to investigate the Planck scale have so far been guided by the belief that nature is fundamentally quantum mechanical, which ultimately rests on Bell-type theorems and related experiments \cite{Clauser1969}. These demonstrate that non-local effects related to measuring quantum particles cannot be manifestations of a deeper, hidden variable layer that obeys locality, as well as a form of statistical independence. The latter is a subtle, albeit crucial mathematical assumption required by proofs of Bell-type theorems, which has never been separately verified. This assumption has recently been called into question increasingly often \cite{tHooft2007, Vervoort2013, Hossenfelder2020, Andreoletti2022}, and could be the main reason why we do not yet understand the Planck scale \cite{Hance2022}. The time has come to systematically investigate the possibility that, fundamentally, nature violates statistical independence, but satisfies locality. While relevant research is still in its infancy, the automata paradigm \cite{tHooft2016} provides a promising, though speculative avenue to physical theories based on these principles.

The automata paradigm argues that, at the Planck scale, physics reduces to finite amounts of data being processed in discrete steps by discrete automata – elementary computing entities that store, process and exchange data according to predefined rules. While individual automata might obey simple rules, large systems of interconnected automata could exhibit very rich behavior: there might exist a multi-automaton model capable of explaining all empirically verified aspects of quantum theory and general relativity. While remarkable correspondences have been demonstrated between simple, deterministic automata and stylized quantum models \cite{tHooft1988, tHooft2013, tHooft2016, tHooft2021}, establishing a connection to general relativity remains very challenging: the latter uses a continuous spacetime, which exhibits local Lorentz invariance (symmetry under space rotations and spacetime boosts), gravitational distortions induced by matter/energy, and an underlying expansion of the entire Universe; moreover, everything is locally-defined, including the flow of time. This is at odds with spatial (cellular) automaton models \cite{Hoekstra2010}, where space is a fixed, regular grid of cells, synchronously updated at each step, using a global, discrete notion of time. This conflict appears to hinder a wider engagement of the scientific community with the automata paradigm. However, it is a conflict that I believe we can resolve in a rather graceful manner, by following a research path that I explain below.

\begin{figure*}
\includegraphics[width=2.0\columnwidth]{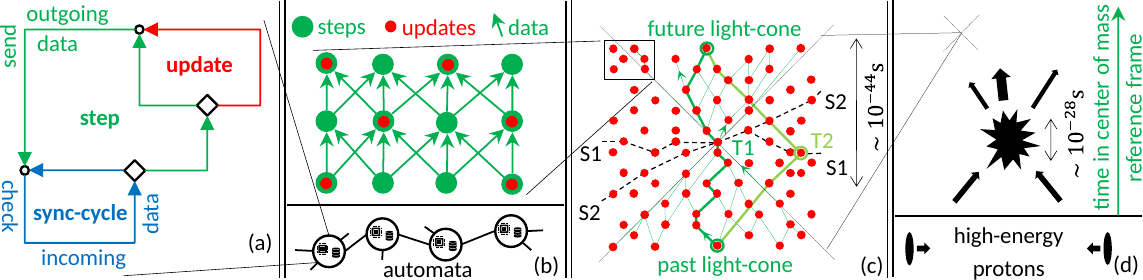}
\caption{\label{fig:wide}General intuition. Outline of envisioned, automaton-level algorithm (a); several, connected automata taking steps and experiencing updates (b), which contribute to a larger DAG of updates, locally approximated by a 2-dimensional spacetime slice (c), which is a tiny part of the background of a proton-proton collision (d). Execution is highly asynchronous (any automaton may take as many steps as allowed by the availability of data expected from neighbors), so not bound to a particular spacetime foliation, like one induced by spatial slice S1 or S2 (c). The number of DAG links along an update chain determines the physical time elapsed along a compatible continuum trajectory, with non-inertial trajectories experiencing less time (T2 compared to T1, illustrating the twin effect) due to there being fewer links available for compatible chains (c). For visual clarity, I only show DAG links in the past and future of one update (with the exception of T2) that cannot be omitted due to transitivity (c).}
\end{figure*}

Since the {\it locality principle} is a central motivation of the automata paradigm, it is crucial that it is strictly satisfied. Specifically, an automaton must only use data it already stores or it has just received from adjacent automata. Any change occurring in the system directly follows from local, automaton-level computation. Importantly, this includes changes of the adjacency structure (the network of connections between automata): creating a link between two, previously disconnected automata requires that some form of token produced by one has reached the other, after traveling over already existing links. There can be no reference to any global entity or absolute time variable, that would perform external interventions/modifications or facilitate any form of centralized coordination. Synchronization between automata must take place locally, by means of signals/messages exchanged between adjacent automata: each automaton cyclically checks that it has received a new (often empty) data packet from each neighbor, before proceeding with the next step of the computation, which is concluded after sending data packets to all neighbors (Fig.~1a). The frequency of this cyclic checking (essentially a refresh rate encoding the number of cycles between two consecutive steps) is meant to be arbitrary: it may change across automata and along the computational thread of any automaton in any way, without any consequence for the system's evolution or the physics describing the system's macroscopic behaviour (we come back to this point below).

These specifications induce a directed acyclic graph (DAG) of steps interconnected by data-dependence links (Fig.~1b). The step-DAG already exhibits, at least locally, a causal structure somewhat similar to the light-cone structure of relativistic spacetime. However, spacetime cannot be an immediate, macroscopic manifestation of the step-DAG, because of the regularity that the latter exhibits: for any pair of steps that are causally connected related, the number of intermediate steps is path-independent, so cannot constitute a microscopic counterpart of relativistic proper time, which is path-dependent. In order to overcome this problem, I propose a distinction between updating and non-updating steps (Figs. 1a, 1b): the former entail some data modification before sending data packets, while the latter only forward incoming data to adjacent automata (the notions of ``data'' and ``data packets'' are further explained below). This should go hand in hand with requiring that modifications of the adjacency structure may only take place during {\it updates} (updating steps). An update may then be understood as an interaction between incoming data, stored data and/or the local network topology. It allows the automaton to write and modify data, in addition to reading and moving data, which may occur during any step. Thus, updates may be irreversible, and it becomes very sensible to use the number of updates along a path as a counterpart of proper time, since it essentially measures the local amount of change in the system.

Provided that it satisfies certain properties (on which I elaborate below), the resulting DAG of data dependence between updates (Fig.~1c) is a sensible way of conceptualizing relativistic spacetime at a microscopic level, in a manner compatible with the path-dependent nature of proper time (which is directly related to the twin effect, illustrated in Fig.~1c). Any type of spacetime (Lorentzian manifold) compatible with general relativity can actually be discretized as a DAG (at least if the manifold lacks causal loops), an insight that has already been extensively used by the causal sets paradigm \cite{Surya2019, Dowker2013}, which aims for a quantum-theoretic description of the Planck scale. DAGs provide a natural way of encoding relativistic spacetime, as well as a natural way of mapping out distributed computation~\cite{Cosnard1999} (as illustrated at above), thus providing the perfect conceptual bridge between the two. I use this bridge to argue that spacetime could be a by-product of distributed computation executed by a giant multi-automaton system, where fundamental events composing the former correspond to local updates (in the specific sense defined above) experienced by the latter, and DAG links encode how events/updates may influence each other and must thus precede each other.

Updates that are not connected by a directional causal chain correspond to spatially-separated events, which cannot influence each other. This enables a discrete counterpart of ``relativity of simultaneity'', and Fig.~1b illustrates how one may choose multiple spatial (simultaneity) slices that go through any update. These slices induce different foliations of the update-DAG, which correspond to different coordinate systems that one may define on the spacetime manifold. Once fully defined, the metric structure inherent to such a manifold is independent of coordinate system choices (spacetime intervals between events are invariant), which can be fully explained by the topology of the update-DAG being independent of conventions that one may use for labeling the updates. But general relativity involves a type of {\it invariance} that is stronger than that: Einstein's field equations, which dynamically determine the manifold's metric structure, are also invariant under coordinate transformations. This implies that the evolution of the multi-automaton system must be strictly independent of any ordering of steps that are not tied by communication-and-synchronization chains (including data-independent updates). Thus, an explicit simulation should be able to execute independent steps (and updates) in any order, or in parallel, while reaching the same, fixed outcome. In the microscopic, multi-automaton description, I refer to this requirement as the {\it invariance principle}, which should be understood as a discrete, computational counterpart of {\it general covariance}.

A crucial challenge is that of jointly and strictly enforcing the locality and invariance principles, both of which are essential for general relativity, and also very meaningful in the context of distributed computation. If these principles are not strictly enforced at a microscopic level, it is highly implausible that they would hold at a macroscopic level, at least under the assumption that the Universe is fundamentally a multi-automaton system. If the automata adjacency structure takes the form of a fixed lattice or network, the cyclic-checking, synchronization mechanism described above (shown in blue in Fig.~1a) provides an immediate solution for enforcing invariance on top of locality. But a fixed automata network is incompatible with Universe expansion and other spacetime curvature effects. Network dynamics is indispensible to the system's evolution, and is the main complication to combining invariance and locality. For instance, a local mechanism for creating new connections between automata can easily break invariance, due to steps and updates that automata experience before becoming connected, when they do not yet synchronize with each other directly. Gracefully handling this type of tension is difficult, and it is not obvious that a rigorous algorithmic solution exists. A system that overcomes this challenge can be said to exhibit {\it relativistic viability}: general relativity would have a reasonable chance of agreeing with the macroscopic behaviour of the system, given that there is an a-priori compatibility in terms of essential principles. 

As a pragmatic way of tackling this challenge, we should aim to construct a network-based protocol governing the communication and rewiring of automata, which would expand the algorithm sketched out in Fig.~1a, in a manner that consistently integrates network evolution operations. The protocol would act as a computational framework/backbone (and a generic set of rules) defining an entire class of viable models, where each model precisely specifies the structure and dynamics of the system in a unique way. For instance, the protocol could handle the evolution of the automata network via link destruction, link creation and node creation operations, while defining protocol-specific data exchange patterns arising when these operations are triggered. A specific model would then define the automaton states that trigger these operations during automaton updates, in terms of additional, model-specific data present at the automaton. It would also specify the automaton states that trigger updates in the first place, the formats governing the model-specific data that is stored and exchanged, and how this data is created. Schematically, a model would specify the dynamics taking place on the network, while the protocol would specify the dynamics of the network, with both model and the protocol playing well-defined roles for specifying how the on-network dynamic drives the of-network dynamic. At this point, it is also worth noting that a data packet sent from one automaton to another may be conceptualized as holding two compartments, for protocol-specific and model-specific data respectively, either of which would often be empty. Even if both compartments are empty, the data packet still serves as a signal used by the synchronization-cycle of the receiving automaton. Only the content of the model-specific data would be forwarded (without modifications) by the receiving automaton during a non-updating step. 

If a protocol with the specifications above can be defined, it can be used to construct viable models, each of which can be explicitly simulated and analysed. This allows us to a-posteriori establish the relativistic and physical correctness of each model, in terms of a multitude of criteria. These falsification criteria can be used in layers: assuming an ordering of criteria in terms of priority, the $k$th criterion would be used on all viable models that have been found to satisfy the previous $k-1$ criteria. The number of models still being considered can be expected to substantially decrease with each criterion, since viable models are unlikely to allow for analytic, a-priori control or predictability of macroscopic, physically interesting quantities, like geometric or even topological properties of the update-DAG. The latter properties should be used for the highest-priority criteria.

Specifically, one should initially check whether the model is capable of generating update-DAGs that converge to Lorentzian manifolds, and in particular to Lorentzian manifolds that are compatible with our Universe. For this purpose, one should first use tools (including dimensionality estimators) already developed \cite{Eichhorn2014, Glasser2013, Reid2003} for the causal sets paradigm. The update-DAG needs to exhibit enough regularity in terms of its local dimensionality, so that a meaningful spacetime embedding exists. Moreover, the measured dimensionality should approach four as the granularity decreases. One should then check that a locally-Minkowski metric, characterized by Lorentz symmetry, emerges at low-enough granularity. This requires that the update-DAG exhibits a high level of randomness, in terms of how updates are distributed within the best-fitting spacetime manifold. Specifically, this distribution should resemble a random sprinkling of a Lorentzian manifold -- that is how physical DAGs are often generated, for illustrative and estimator-validation purposes, in the context of research on causal sets. Note that my formulation effectively constrains updates to a discrete, regular lattice of steps (Figs.~1b, 1c). Even if updates are randomly distributed over the lattice, Lorentz symmetry becomes increasingly broken as the update density $\rho$ (ratio between the number of updates and steps) increases, and the update-DAG becomes increasingly similar to the step-DAG. Thus, a correct model should induce a very small $\rho$ (much smaller than that inherent to Figs.~1b, 1c), in order to satisfy this Lorentzian criterion -- which may be formalized in terms of proper time discrepancies between inertial and non-inertial trajectories with a common start and end (like T1 and T2 in Fig.~1c). This line of reasoning also suggests that one could use high-precision measurements of time-dilation (or other relativistic effects connected to Lorentz symmetry) to search for empirical signatures of the underlying step-DAG, in the form of subtle deviations from relativistic predictions. The magnitude of these deviations would decrease with decreasing $\rho$, which can also be understood as a ratio between two levels of granularity that are both above the Planck scale. For a fixed $\rho$, the magnitude of such deviations should also decrease with an increasing fraction of updates that trigger modifications of the automata network, with an increasing dimensionality (small, compact extra-dimensions could also contribute), and with an increasing amount of clustering of the update-DAG (over the step-DAG).

More advanced falsification criteria should make use of discrete counterparts of the precise, field-theoretic formulations of general relativity and quantum theory. While the relativistic side would deal with macro-distortions of the emergent spacetime and their relationships to matter-energy distributions, the quantum side would deal with micro-regularities/structure existing on top of the emergent spacetime. Such considerations might only become meaningful at granularities substantially lower than those for which notions of dimensionality and Lorentz symmetry become meaningful, so one might need to work with an aggregate variant of the update-DAG (which becomes particularly sensible if, as already mentioned above, the update-DAG shows significant clustering). On the relativistic side, discrete counterparts of curvature-related notions in differential geometry~\cite{Hoorn2023} would play an essential role. On the quantum side, our extensive knowledge of the standard model of particle and interactions, in terms of particle-field content and coupling parameters, should in principle provide substantial falsification power, although there are significant methodological and conceptual challenges when it comes to mapping a complicated quantum field-theoretic system to a multi-automaton system. Micro-structure could be first explored with pattern detection techniques used for studying complex systems \cite{Squartini2017, Babeanu2021}, which allow for the identification of significant deviations (including state space attractors) from meaningful (random matrix/graph) statistical ensembles acting as null models (accounting for the previously-identified, emergent geometry in a maximally-random way). Building on core ideas from earlier work on the automata paradigm, one should aim to understand quantum states as linear combinations of multi-automaton states that a given part of the system may attain, which allows for mapping quantum harmonic oscillators (which are essential to quantum field theories) to state space cycles (which often arise in deterministic, automata-based systems). Disruptions of these cycles, which may be related to interactions between adjacent parts of the system or to using aggregate degrees of freedom for describing the system, are already known to be essential for automata-based interpretations of more complicated quantum systems~\cite{tHooft2020, tHooft2021}. It might be possible to feed multi-automaton state data extracted from simulations to some form of machine learning algorithm designed to infer effective, aggregate time-evolution operators characterizing small regions of the system, which would allow for the extraction of effective Hamiltonians and energy levels. Last but not least, concepts related to string theory could prove valuable for deriving local, quantum-field theoretic descriptions of the multi-automaton system: the update-DAG might exhibit small, compact extra-dimensions, with an associated topology (which would form spontaneously and likely fluctuate, rather than being postulated and fixed) that could be identified and used to infer particle-field contents and associated coupling constants. 

While a multitude of falsification criteria can be defined, at least in principle, using the above considerations, we may still wonder about the number of viable models that one would start with: if this number is very high, it might be unfeasible to apply even the highest-priority criterion to all of them, and one might still be left with an undesirably large number of models even after applying several criteria. It is difficult to say much about the number of viable models before having a constructed a protocol that guarantees relativistic viability, according to specifications above. However, we can expect that a very small subset of the models allowed by the protocol would comply with strict determinism: it is difficult to formulate a consistent set of automaton-level rules that make no use random number generation for tie-breaking purposes, especially when the underlying topology is irregular and dynamic. But determinism lies at the heart of the automata paradigm: like locality, determinism is essential to general relativity, and also important for quantum theory. According to the established understanding, violations of locality and determinism go hand in hand, and fundamentally occur only during instances of quantum measurement, which conceptually is not well understood yet. So it is very sensible that, after being strict about locality, we should also be strict about determinism. Historically, determinism was a core motivation for the discussion that led to the development of Bell-type theorems and experiments, and is essential to violations of statistical independence (allowing for Bell-type theorems to be circumvented) that an automata-driven Universe is expected to exhibit (under the assumption that any event in the past light-cone of any experiment that we can perform is also in the future light-cone of at least one, common event in the early Universe; this is fully compatible with modern cosmology, at least if we admit the existence of an early phase of accelerated expansion capable of solving the horizon problem). In combination with discreteness, determinism is what facilitates the existence of state space cycles, and their mapping to quantum harmonic oscillators (as mentioned above). Finally, in combination with irreversible updates, determinism leads to convergence in state space (self-organization). This implies that desirable properties that the system might exhibit after long-term evolution (like an emergent geometry and micro-regularities that may be described using quantum field theories) could be highly independent of the initial state (which, for simulation purposes, would be randomly sampled from a uniform probability distribution of possible states), thus reducing or entirely eliminating the need of fine-tuning the initial state, which enhances the predictive power (and falsifiability) of viable models, and the multi-automaton approach as a whole. It is this type of state space convergence that I expect would drive an early phase of accelerated expansion, by means of a sudden shift from highly-connected, small-world automata networks, which are very likely to arise (due to statistical, combinatorial reasons) when the initial state is generated in a maximally-random way, to geometric, quasi-random automata networks, characterized by much larger distances. Unlike well-known inflation models, this type of accelerated expansion would go hand in hand with a drastic change in topology (and associated decrease of effective dimensionality). This would essentially be a form of geometrogenesis~\cite{Konopka2008, Bianchi2023} occurring in real time, thus circumventing conceptual issues pertaining to atemporality or timelessness, as well as the need of suitably changing a global temperature parameter that would control a predefined statistical ensemble of network configurations.

While the number of viable models might remain high even after invoking the requirement of strict determinism, it is very likely that remaining models would be easily differentiated in terms of their (a-priori) simplicity, with a very small number of very simple models, and increasing numbers of models available at decreasing levels of simplicity. This simplicity should be primarily quantified in terms of the format governing stored and exchanged, model-specific data, as well as the complexity of automaton-level, model-specific rules driving the computation. Practical and predictive power considerations allow us to give priority to the simplest possible models, and to progressively increase the algorithmic and data complexity. They also allow as to ignore, at least initially, models that require free parameters -- especially since we would not even control the dimensionality of the system, which is effectively a free (positive integer) parameter in general relativity, which otherwise has very few parameters.

The multi-automaton formulation seems to allow for an even stronger type of falsification, by predicting the model itself, based on evolutionary (natural selection) arguments, while using little or no anthropic (observation selection) arguments. Specifically, one could argue that all/many viable models might have jointly governed the system in the distant past, with different automata operating under different models, all complying with the same communication-and-rewiring protocol. The protocol needs to provide a local, automaton-creation operation (for consistency with the well-established, dimensionality-conserving expansion of the Universe), so a newborn automaton would need to ``inherit'' the model of its parent automaton. This leads to competition between automaton ``species'', with species more efficient at reproducing being ``naturally selected''. Such an {\it automatogenesis} hypothesis could be examined via computer simulations, which might make predictions about a winning model (or combination of symbiotic models). Simpler models might even have an evolutionary advantage over more complicated ones, given that rules of the former (including those governing automata creation) would require simpler inputs, and could thus make use of (truncated versions of) incoming data produced by the latter, but not the other way around. Obviously, this reasoning assumes the existence of a shared, underlying protocol. While it might also be possible to use such reasoning to explain how the protocol itself might have been selected, it is much more difficult. Still, a relevant observation can be made, in relation to the local synchronization mechanism described above, which would arguably the most fundamental part of the envisioned protocol. Specifically, regions of the multi-automaton system that violate synchronization (where automata do not always wait for data packets to arrive from all neighbors before moving to the next step) would exhibit a fundamental form of indeterminism. Locally, this allows for all kinds of states which otherwise would not have been possible (at least conditionally on states in the past light-cone), which compromises regularities, cycles and rich behaviour that the system might otherwise exhibit on the long-term -- in the context of cellular automata, the crucial role that synchronization plays for pattern formation has been known for a long time~\cite{Bersini1994}. This might affect the ability of non-synchronizing automata to reproduce, remain connected or effectively influence the rest of the system, but is something that would require extensive investigations in a dynamically-networked context. Still, synchronization violations could prevent the structure (including geometry) characterizing our Universe to arise, along with any form of memory or intelligence, which at least allows for an anthropic argument to be constructed (one that is rather strong, perhaps unavoidable and even desirable at this very fundamental level).

Motivated by relativistic aspects, the possibility of using random-geometric DAGs in the context of the automata paradigm had already been mentioned in Ref.~\cite{tHooft2016-c}, but without further elaborations on how the two could be sensibly integrated. Until now, progress on this matter appears to have been hindered by the non-locality of existing DAG generating procedures used for research on causal sets. For instance, the classical sequential growth approach~\cite{Rideout1999} relies on global access to all the previously generated spacetime events, upon randomly generating causal links to every new event. This type of non-locality is a widespread feature of random graph generation algorithms -- like the preferential attachment model~\cite{Barabasi1999} -- regardless of the properties they enforce (so not confined to geometric random graph generation). Probabilistic Markov Chain procedures for sampling graphs from a predefined statistical ensemble also assume access to the entire graph upon randomly picking the location where a modification at a given step is attempted, despite the modification usually being small, highly localized. Going back to discrete spacetimes, deterministic DAG-generation procedures that have been previously proposed~\cite{Wolfram2002, Bolognesi2011, Wolfram2020, Bolognesi2017} suffer from similar issues, involving external entities that sequentially perform small, localized modifications of a discrete system to which they have global access. Despite their limitations, such procedures may provide important insights about the types of network motifs, micro-structures and localized modifications that facilitate geometric emergence~\cite{Chen2013}, with the approach in Ref.~\cite{Akara-pipattana2021} being of particular value, thanks to its maximum-entropy ensemble formulation, which minimizes bias and the possibility of introducing unconscious assumptions. Also note that Ref.~\cite{Wolfram2002} mentions, in the context of cellular automata with fixed topology, a signal-exchange idea similar to the local synchronization mechanism proposed here, but does not use it in the context of graph rewriting models proposed for describing the Universe. It also discusses a {\it causal invariance} property that DAG generation from graph rewriting would need to exhibit, which is very similar to the invariance principle here, but it appears very difficult (perhaps impossible) to formulate a physically sensible graph rewriting procedure that would strictly observe it. 

Note that the protocol-model distinction that I used here is somewhat similar to that used in~\cite{Babeanu2023}, in the context of adaptive (shared-memory) parallelism of multi-agent-based simulations. While the protocol developed there enforces a counterpart of invariance, it uses global entities and only provides a way of running parallel simulation of any model (which is advantageous if the model involves localized updates), instead of a way of constructing models with specific properties. The protocol that I anticipate here might also be relevant for the purpose of handling (distributed-memory) parallelism of scientific simulations, in a context where message passing takes place on top of a hardware topology that changes during the simulation and lacks centralized control. Despite such connections, this work is not driven by an underlying belief or intuition that our Universe is a computer simulation. On the contrary, the strict notion of locality advocated here would be of little use for mathematically describing a simulated Universe, which would anyway involve external entities (like central processing units, simulation managers and researchers) with instantaneous access to arbitrarily large parts of the simulated system. The underlying intuition is that our Universe is real, has a well-defined, discrete micro-structure and is governed by algorithmic laws of classical computation. 

\vspace{1cm}

\begin{acknowledgments}
I acknowledge inspiring discussions with Michael Thompson and Marco Verweij on fundamental problems in social science, which were the initial triggers for my interest in self-organization and dynamically-networked, multi-agent/automaton systems. While becoming increasingly aware of the relevance these notions have for fundamental physics and computer science questions, I also benefited from interactions with Gerard Barkema, Wendelin B\"{o}hmer, Oleg Evnin, Diego Garlaschelli, Betti Hartmann, Yaroslav Herasymenko, Hildegard Meyer-Ortmanns, Tim Petoukhov, Felix Reed-Tsochas, Jason Roos and Neil Yorke-Smith. In addition, I acknowledge feedback from Iacopo Bertelli, Maaike Elgersma, Eugenia Rosu, Andra Stroe, Pascal van der Vaart, Bas Verbruggen and ten anonymous readers, in relation to a research proposal from which this manuscript developed. The work was was partially supported by the TU Delft Institute for Computational Science and Engineering, by TAILOR, a project funded by the EU Horizon 2020 research and innovation programme under grant~952215, and by the Netherlands Organization for Scientific Research (NWO/OCW) via grant~314-99-400.     
\end{acknowledgments}

%\appendix

%\section{Appendix Section}

\bibliography{paper}% Produces the bibliography via BibTeX.

\end{document}